# Joint Temporal-Structural Representation Learning for Distributed Fault Discrimination in Microservice Architectures

Yihan Xue
University of Southern California
Los Angeles, USA

Yuxiao Wang
University of Pennsylvania
Philadelphia, USA

Ao Zhu
University of Pennsylvania
Philadelphia, USA

Xiaoxuan Sun
Independent Researcher
Mountain View, USA

Chong Zhang*
Carnegie Mellon University
Pittsburgh, USA

***Abstract*-Addressing the diverse fault morphologies, complex dependencies, and time-varying operational states in microservice distributed systems, this paper proposes a distributed fault discrimination model based on temporal graph neural networks. This model characterizes the microservice operation process as a dynamic graph sequence evolving, and performs joint representation learning of temporal modeling and structural interactions within a unified framework. First, service-level multi-source observation signals are aligned and characterized to construct node feature sequences and their corresponding time-dependent dependencies. Then, a temporal coding module is introduced to extract the dynamic evolution representation of service states, and at each time step, attention-based structured message passing is used to characterize dependency interactions and propagation associations, forming a structure-enhanced temporal node representation. Furthermore, a dual readout mechanism is employed to aggregate the node and temporal dimensions, obtaining a system-level global representation and outputting the fault category distribution. Finally, supervised learning objectives are used to optimize model parameters, enabling the model to learn stable discrimination evidence under complex interactions and multi-source noise conditions. Comparative experimental results show that the proposed method achieves superior performance on multiple evaluation metrics, validating the effectiveness of jointly modeling temporal evolution and dependency structures in improving the distributed fault discrimination capability of microservices.**

## I. INTRODUCTION

Cloud-native architecture has driven the large-scale deployment of microservices in the internet and enterprise systems. Business capabilities are broken down into numerous autonomous service units, forming complex dependency networks through call chains and messaging mechanisms. This architecture improves iteration efficiency and elastic scaling capabilities, but also introduces higher runtime uncertainty and fault propagation risks. Service dependencies are dynamically changing; resource contention, rate limiting, circuit breaking, version rolling, and instance migration can lead to faults exhibiting cross-service propagation, time-varying coupling, and multi-source heterogeneity[1]. Traditional diagnostic strategies relying on single-point metrics or static topology struggle to characterize these complex relationships, easily leading to unclear identification and location errors in scenarios with multiple concurrent or cascading faults, thus affecting system stability and operational efficiency[2].

The core challenge of distributed fault diagnosis lies in simultaneously understanding temporal evolution and structural dependencies. The observational data generated by microservice systems encompasses multimodal information such as metrics, logs, and traces, characterized by high dimensionality, noise, missing data, and misalignment. Furthermore, the signal correlations between different services are often influenced by both call topology and runtime load. Faults are not merely local anomalies, but rather accompany delays, amplification, or obscuration along dependent links, forming structured patterns that progress over time. Modeling only in the time dimension makes it difficult to distinguish between global fluctuations caused by external pressures and propagation effects triggered by local anomalies. Modeling only in the graph structure dimension makes it difficult to capture state transitions and causal temporal characteristics before and after a fault. Therefore, unifying temporal information and graph structure into a single discriminative framework is a key path to improving the robustness and interpretability of distributed fault detection[3].

The modeling approach based on temporal graph neural networks can jointly learn the evolution of node states over time and the structural interactions of service dependencies on a dynamic graph, providing a more mechanistic representation for microservice fault detection[4]. By extracting short-term mutation and long-term drift features in the time dimension and aggregating neighborhood dependencies and cross-layer propagation information in the graph space, a discriminative representation that considers both local anomalies and global impacts can be formed, helping to mitigate uncertainties caused by noise interference and incomplete observations. Meanwhile, sequence graph modeling can naturally incorporate structural priors such as call relationships, link strength, and interaction direction, elevating faults from a single-service perspective to a system-level perspective, enabling more refined differentiation between cascading faults and multi-point common-cause scenarios. This type of model can also provide structured explanatory clues through attention weights, information flow paths, and subgraph contributions, providing traceable evidence for operational decisions.

## II. METHODOLOGY FOUNDATION

Microservice The methodological foundation of this work is grounded in the convergence of spatiotemporal representation learning,

graph-based dependency modeling, and robust learning under complex data conditions, all of which collectively inform the design of a temporal graph neural architecture for distributed fault discrimination in microservice systems.

A primary methodological pillar is the use of graph-based modeling to capture inter-service dependencies, where service interactions are naturally represented as structured relational data. Prior work has demonstrated that graph neural networks can effectively model anomalous propagation patterns in distributed microservices by encoding call relationships and traffic dependencies into node embeddings [5]. However, such approaches largely rely on static or weakly dynamic structures, limiting their ability to capture evolving system states. This limitation motivates the transition toward dynamic graph representations, where both topology-aware interactions and temporal evolution are jointly modeled. Extending this direction, recent advances in spatiotemporal representation learning integrate graph structures with sequence modeling mechanisms, enabling the capture of both structural correlations and temporal transitions in complex systems [6]. These studies provide a direct methodological basis for constructing dynamic graph sequences in which node states evolve over time while preserving dependency-aware message passing.

Complementing graph-based modeling, sequential learning frameworks play a crucial role in capturing temporal dynamics inherent in microservice operations. Data-centric sequential modeling paradigms emphasize the importance of aligning heterogeneous time-series observations into unified representations, thereby improving decision intelligence at the session level [7]. Similarly, sequence modeling techniques tailored for cloud-native fault detection introduce cost-sensitive objectives and advanced sequence encoders to better capture long-term dependencies and rare fault patterns [8]. These approaches highlight the necessity of integrating temporal encoding mechanisms capable of modeling both short-term fluctuations and long-term drifts, which directly informs the temporal coding module in the proposed framework. Furthermore, uncertainty-aware time series forecasting methods demonstrate that modeling prediction uncertainty can enhance robustness against noisy and volatile backend metrics [9], reinforcing the need for stable temporal feature extraction under real-world operational noise.

Another critical methodological dimension involves robust learning under data heterogeneity, imbalance, and distribution shift, which are prevalent in microservice environments. Self-supervised learning strategies have been shown to enhance fault diagnosis performance under severe class imbalance by leveraging intrinsic data structures to learn invariant representations [10]. Similarly, meta-learning approaches enable rapid adaptation to few-shot or rare-event scenarios, improving generalization in sparse fault conditions [11]. Methods addressing joint class imbalance and distribution shift further demonstrate the importance of designing learning objectives that remain stable across varying data distributions [12]. These works collectively motivate the adoption of supervised objectives augmented with robustness considerations, ensuring that the learned representations remain discriminative despite noise, imbalance, and evolving system behaviors.

In parallel, advances in multi-source data integration and shared representation learning provide essential support for modeling heterogeneous observability signals. Techniques for shared representation learning across high-dimensional and multi-task settings enable effective fusion of diverse data modalities such as metrics, logs, and traces [13]. Generative and reconstruction-based anomaly detection methods further illustrate how multi-scale temporal features can be extracted from complex system signals [14], while distributed tracing-based approaches highlight the importance of fine-grained temporal alignment across service interactions [15]. These insights directly inform the design choice of constructing aligned multi-source node feature sequences, ensuring that both temporal consistency and cross-service comparability are preserved. At the system level, emerging research on intelligent decision systems and agent-based frameworks underscores the value of structured state representation and hierarchical reasoning in complex environments. Structured state modeling and constraint-guided learning approaches demonstrate how high-level system representations can support downstream decision-making tasks [16], while memory-driven and multi-agent frameworks highlight mechanisms for aggregating long-horizon information and coordinating distributed intelligence [17-18]. Additionally, integrating large language models with observability data has shown promise in enhancing automated root cause analysis by leveraging semantic reasoning over system states [19]. These developments collectively inspire the dual readout mechanism in the proposed model, where node-level and temporal-level aggregations are combined to form a global system representation suitable for fault classification and higher-level operational reasoning. Finally, complementary system-oriented works on cloud-native environments, including serverless inference optimization and microservice development frameworks [20-21], provide contextual grounding for scalability and deployment considerations. Although not directly contributing to the core modeling architecture, they reinforce the necessity for efficient, scalable, and interpretable models that can operate in production-grade distributed systems.

# III. METHOD

Given an observation sequence of a microservice system over discrete time steps, a time-evolving directed graph sequence $\{\mathcal{G}_t\}_{t=1}^{T}$ is constructed, where $\mathcal{G}_t = (\mathcal{V}, \mathcal{E}_t)$ denotes the service dependency relations at time step $t$, $\mathcal{V}$ is the set of service nodes and remains invariant within the time window, and $\mathcal{E}_t$ is the time-dependent set of invocation edges. After aligning multi-source observable signals to the service granularity, the input feature of the node $v_i$ at time step $t$ is denoted as $\mathbf{x}_{i,t} \in \mathbb{R}^{d_x}$, and the adjacency matrix is defined as $\mathbf{A}_t \in \{0,1\}^{|\mathcal{V}|\times|\mathcal{V}|}$. The formal definition is given as follows.

$$\mathcal{G}_t = (\mathcal{V}, \mathcal{E}_t),\ \mathbf{A}_t(i,j) = \mathbb{I}\left((v_j \rightarrow v_i) \in \mathcal{E}_t\right),\ \mathbf{x}_{i,t} \in \mathbb{R}^{d_x} \quad (1)$$

In this modeling framework, the edge direction is defined according to the directed relationship from the caller to the callee, which imposes a semantic constraint ensuring that the process of information aggregation remains aligned with the paths through which faults propagate. This directional design is explicitly applied to preserve causality in service invocation chains, thereby enabling the model to leverage structural priors for modeling propagation dynamics. The overall architecture of the proposed model is illustrated in Figure 1.

To construct robust node representations under multi-source and time-varying observations, this work builds upon the sequence modeling paradigm introduced by C. Zhang et al. [22], which models protocol behaviors as ordered status code sequences for anomaly detection. Their method fundamentally captures temporal dependencies by learning discriminative patterns from sequential observations. Inspired by this principle, our framework adopts temporal encoding mechanisms to transform aligned metrics, logs, and trace signals into dynamic node feature sequences, thereby incorporating both short-term fluctuations and long-term evolution trends into the representation space. To further enhance temporal adaptability under dynamic workloads and evolving system states, this study leverages the idea of handling workload drift proposed by W. Huang et al.[23], where reinforcement learning is used to stabilize cloud scaling decisions under non-stationary environments. While their method focuses on decision policies, its core principle of learning under distributional shifts is incorporated into our temporal modeling module by enabling the encoder to adapt to time-varying data distributions, thus improving robustness against concept drift and noisy observations in microservice environments. In modeling inter-service dependencies and resource interactions, this work builds upon the scheduling and resource coordination strategies proposed by X. Li et al [24]. in DeepServe, which jointly considers service-level objectives and cost constraints for multi-tenant inference systems. Their method fundamentally models dependencies between tasks and resources to optimize global system performance. Drawing from this idea, our approach adopts dependency-aware representation learning, where service nodes are not treated independently but are incorporated into a unified graph structure, allowing the model to leverage interaction-aware message passing for capturing fault propagation across services.

Furthermore, to refine structural interaction modeling at a finer granularity, this study leverages the token-level resource coordination mechanism introduced by K. Zeng et al [25]. in TokenFlow. Their method decomposes workloads into fine-grained units and dynamically schedules them to optimize concurrency. Analogously, our model adopts fine-grained attention-based message passing, where node interactions are dynamically weighted at each time step, enabling the framework to capture heterogeneous dependency strengths and selective propagation paths, thereby extending coarse-grained graph aggregation into a more adaptive and context-aware mechanism. To enhance global coordination and system-level optimization capability, this work incorporates the multi-agent interaction modeling paradigm proposed by C. Wang et al. [26], which combines deep reinforcement learning with game-theoretic structures to model cooperative and competitive behaviors among agents. The fundamental contribution of their method lies in capturing interdependent decision-making processes. Building on this principle, our framework models services as interdependent entities within a graph and leverages attention mechanisms to approximate coordinated interactions, thereby extending the concept of multi-agent coordination into a differentiable representation learning process for fault discrimination.

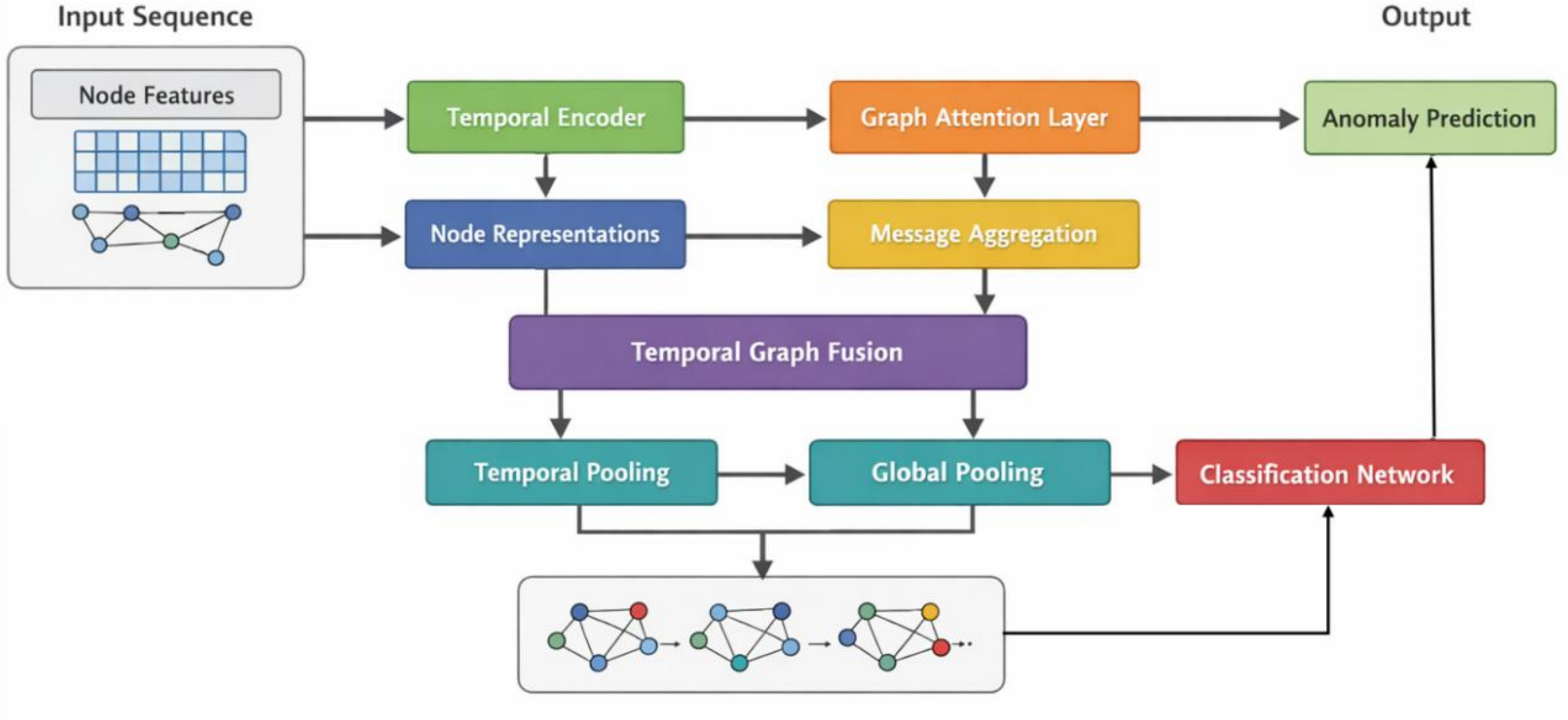

Figure 1. Overall model architecture

To capture the temporal evolution of service states, a temporal encoder with shared parameters is introduced to map $\mathbf{x}_{i,t}$ to a hidden state $\mathbf{h}_{i,t} \in \mathbb{R}^{d_h}$. Specifically, the input is first linearly projected to obtain $\mathbf{u}_{i,t}$, and then a gated recurrent unit is used to recursively update historical states, so that short-term perturbations and long-term drift information are encoded jointly in the same latent space.

$$\boldsymbol{u}_{i,t} = \boldsymbol{W}_x \boldsymbol{x}_{i,t} + \boldsymbol{b}_x \quad (2)$$

$$\boldsymbol{h}_{i,t} = GRU\left(\boldsymbol{u}_{i,t}, \boldsymbol{h}_{i,t-1}\right) \quad (3)$$

This temporal representation provides time context for subsequent graph-structured interactions, enabling structural messages to exhibit different effective strengths and response patterns across time steps.

At each time step $t$, structured message passing is performed based on $\boldsymbol{A}_t$ to characterize the interaction effects induced by service dependencies. For node $v_i$, messages are collected from its in-neighborhood $\mathcal{N}_t(i) = \{j | \boldsymbol{A}_t(i,j) = 1\}$, and an attention mechanism is adopted to adaptively weight dependency strengths from different sources. First, the edge-wise relevance score $e_{ij,t}$ is computed and normalized to obtain the attention weight $\alpha_{ij,t}$, and then weighted aggregation over the neighborhood yields the structural message $\mathbf{m}_{i,t}$.

$$e_{ij,t} = \boldsymbol{a}^{\top} \sigma\left(\boldsymbol{W}_q \boldsymbol{h}_{i,t} + \boldsymbol{W}_k \boldsymbol{h}_{j,t} + \boldsymbol{b}\right) \quad (4)$$

$$\alpha_{ij,t} = \frac{exp\left(e_{ij,t}\right)}{\sum_{k \in \mathcal{N}_t(i)} exp\left(e_{ik,t}\right)} \quad (5)$$

$$\boldsymbol{m}_{i,t} = \sum_{j \in \mathcal{N}_t(i)} \alpha_{ij,t} \, \boldsymbol{W}_v \boldsymbol{h}_{j,t} \quad (6)$$

Finally, the temporal hidden state and the structural message are fused and updated to obtain the temporal graph representation $\boldsymbol{z}_{i,t}$, which can be regarded as a structure-enhanced state under temporal context constraints.

$$\boldsymbol{z}_{i,t} = \sigma\left(\boldsymbol{W}_s \left[\boldsymbol{h}_{i,t} \parallel \boldsymbol{m}_{i,t}\right] + \boldsymbol{b}_s\right) \quad (7)$$

To enable distributed fault diagnosis, node representations within the time window are aggregated into a system-level representation and mapped to a fault-category distribution. A dual readout strategy over time and node dimensions is adopted: node pooling is first applied at each time step to obtain $\boldsymbol{g}_t$, then temporal aggregation is performed to form the global representation $\mathbf{g}$, and finally, a classifier outputs the predicted probability $\widehat{\boldsymbol{y}}$. Meanwhile, a supervised objective is used to constrain the model to learn discriminative system-level feature representations.

$$\boldsymbol{g}_t = Pool\left(\{\boldsymbol{z}_{i,t}\}_{i \in \mathcal{V}}\right), \quad \boldsymbol{g} = Pool\left(\{\boldsymbol{g}_t\}_{t=1}^{T}\right) \quad (8)$$

$$\widehat{\boldsymbol{y}} = softmax\left(\boldsymbol{W}_c \boldsymbol{g} + \boldsymbol{b}_c\right) \quad (9)$$

$$\mathcal{L} = -\sum_{c=1}^{C} y_c \, log \, \widehat{y}_c \quad (10)$$

Here $C$ denotes the number of fault categories, and $Pool(\cdot)$ can be instantiated as mean pooling or attention pooling to accommodate the representation requirements of systems with different scales, thereby completing distributed fault diagnosis modeling without relying on a fixed topology size.

## IV. Experimental Results and Analysis

### A. Dataset

This study adopts the GAIA dataset as an open-source benchmark for microservice distributed fault diagnosis. Designed for intelligent operation and maintenance scenarios, GAIA integrates multi-source observational data together with controlled fault injection records in microservice environments. Such characteristics make the dataset well suited for constructing discriminative models that capture both temporal evolution patterns and inter-service dependency relationships. The dataset is primarily derived from the MicroSS microservice business simulation system, which continuously gathers metrics, logs, and distributed tracing information during system operation. In addition, the platform records injected anomaly events and associates them with specific services and timestamps. These annotations provide reliable supervision signals and support the study of distributed fault diagnosis under conditions involving dynamic service topology and multimodal monitoring signals.

From the perspective of data organization, the MicroSS subset is structured into several main directories, including metric, trace, log, and runtime records. The trace files contain fields related to call chains, such as timestamps, service identifiers, trace and span IDs, parent – child relationships, and request status information. These attributes enable the natural construction of time-varying service interaction graphs and facilitate modeling of fault propagation paths. The metric files primarily consist of timestamped numerical measurements, covering a wide range of operational indicators for services and middleware components. The log files store textual system logs that are grouped by service instances. In addition, the run directory provides system execution logs together with anomaly injection records, allowing the alignment of system-level fault labels with service-level observations. This integrated structure supports the consistency requirements of temporal graph neural networks, particularly with respect to node feature representation, evolving graph topology, and supervised learning targets.

### *B. Experimental Results and Analysis*

To comprehensively evaluate the distributed fault discrimination capability of microservices based on temporal graph neural networks, this paper selects several representative methods related to microservice anomaly detection and root cause localization as comparison objects, and reports the performance of each method on multi-dimensional evaluation metrics under a unified data processing and evaluation protocol. The evaluation metrics cover classification discrimination, detection capability, and consistency measurement, so as to characterize the comprehensive discrimination features of the model under complex dependencies and time-varying operating conditions. The experimental results are shown in Table 1.

Table 1. Comparative experimental results

| Method | Accuracy | Precision | Recall | F1 | AUC-ROC | Macro-F1 | Micro-F1 | MCC |
|---|---|---|---|---|---|---|---|---|
| Zhang et al.[27] | 0.82 | 0.82 | 0.79 | 0.79 | 0.85 | 0.78 | 0.81 | 0.65 |
| Chen et al.[28] | 0.84 | 0.82 | 0.81 | 0.81 | 0.86 | 0.79 | 0.83 | 0.67 |
| Lin et al.[29] | 0.85 | 0.83 | 0.82 | 0.82 | 0.87 | 0.8 | 0.84 | 0.68 |
| Gan et al.[30] | 0.86 | 0.84 | 0.83 | 0.83 | 0.88 | 0.81 | 0.85 | 0.69 |
| Zhao et al.[31] | 0.87 | 0.85 | 0.84 | 0.84 | 0.89 | 0.82 | 0.86 | 0.7 |
| Zheng et al.[32] | 0.88 | 0.86 | 0.85 | 0.85 | 0.9 | 0.83 | 0.87 | 0.71 |
| Zhang et al.[33] | 0.89 | 0.87 | 0.86 | 0.86 | 0.91 | 0.84 | 0.88 | 0.72 |
| Ours | 0.93 | 0.92 | 0.91 | 0.92 | 0.95 | 0.9 | 0.93 | 0.78 |

From a multi-dimensional evaluation perspective, different methods show clear performance gaps while remaining consistent across metrics, indicating that failure discrimination quality is jointly reflected in accuracy, detection capability, and consistency. Some approaches exhibit a precision–recall trade-off, reducing false positives at the cost of missing boundary cases or introducing noise when improving coverage, which impacts overall performance. In contrast, macro-F1 is more sensitive to class imbalance and long-tail failures, better capturing robustness on rare cases. The proposed method achieves superior results across all metrics, demonstrating stronger stability and generalization. Its balanced precision and recall lead to consistent overall gains, suggesting more effective joint modeling of failure signals and propagation effects. Additionally, improved consistency indicates more focused outputs with fewer error types, providing a more reliable foundation for fault handling and system-level decision-making.

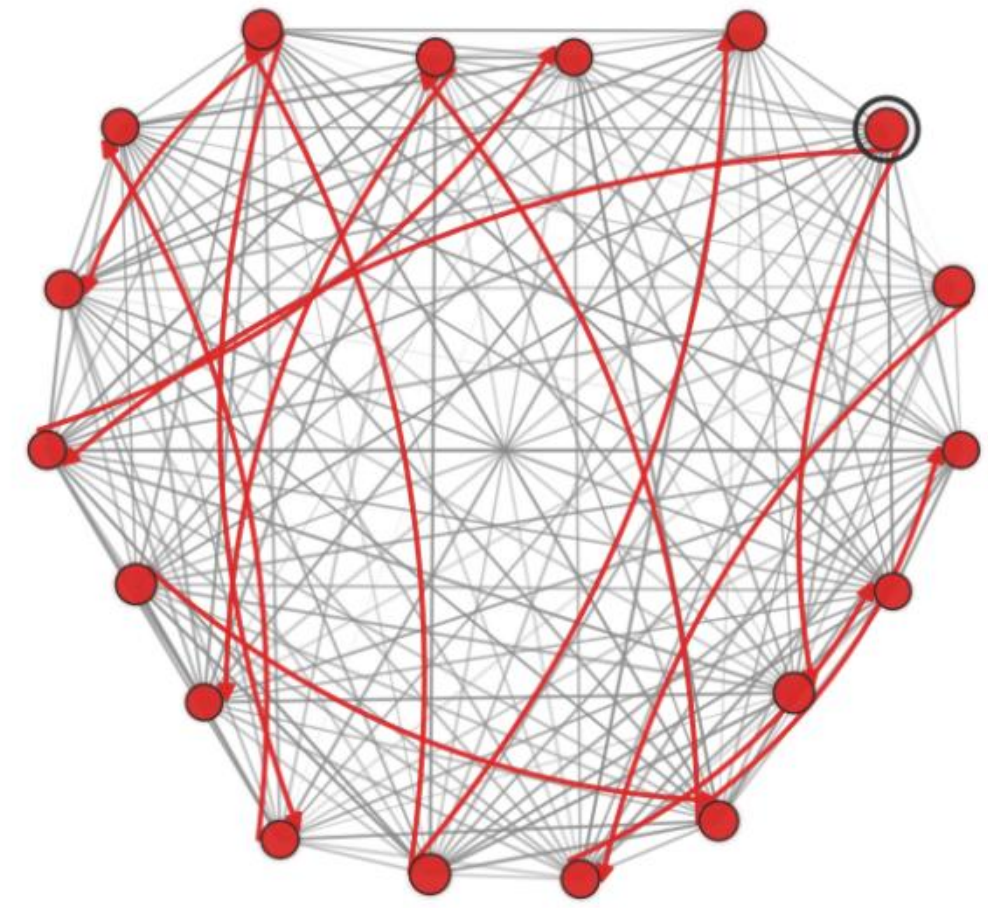

Figure 2. The visualization results of the dynamic call graph's temporal evolution.

Gray edges represent service call relationships aggregated over time, while red nodes and arrows indicate fault impact and propagation paths. As shown in Figure 2, the dynamic call graph reveals a dense dependency network with strong inter-service coupling and multiple multi-hop paths, making it easier for local anomalies to spread and cause system-level effects. The gray edges provide a global view of potential propagation channels, emphasizing the need to consider both local and long-range dependencies in fault diagnosis. The red nodes and arrows illustrate how faults expand along key paths, with hub nodes exhibiting convergence and amplification effects. Overall, this visualization supports time-series graph modeling by highlighting the importance of focusing on propagation-related subgraphs and their temporal context for accurate fault identification.

## V. Conclusion

This paper addresses key challenges in microservice distributed systems, such as difficult fault identification, complex propagation paths, and heterogeneous and time-varying observed signals. It proposes and systematically constructs a distributed fault identification modeling approach based on temporal graph neural networks. This method adopts a unified perspective of system dependency structure and runtime evolution, mapping multi-source observations into dynamic graph sequences and implementing time encoding, structural interaction, and system-level readout within the same framework. This enables the model to characterize fault-related state transitions and cross-service correlations under service dependency constraints. This research strengthens the paradigm shift in microservice fault identification from single-point indicator-driven to joint structural and temporal modeling, providing a more systematic and engineering-applicable theoretical foundation and methodological support for intelligent operation and maintenance in complex cloud-native environments.

From an application impact perspective, for scenarios such as high-concurrency online services, cloud platform infrastructure, distributed middleware, and multi-tenant service orchestration, the proposed framework, with system-level representation at its core, can form a more stable fault identification capability and provide a consistent decision-making basis for operation and maintenance links such as alarm aggregation, impact scope identification, and handling priority ranking. Because this method explicitly incorporates service dependencies and call interactions during the modeling process, it is naturally adaptable to typical production risks such as cascading failures, chain congestion, and cross-service disturbances caused by resource contention. It can more reliably extract key evidence related to failures in complex interaction contexts. Simultaneously, the structured representation and visual representation of propagation paths help reduce the understanding costs associated with black-box discrimination, providing a clearer, institutionalized framework for operational knowledge accumulation, failure review, and experience transfer, thereby promoting the standardization and automation of intelligent operation and maintenance processes.

This work is significant not only for its empirical performance but also for its methodological contribution to distributed systems research. By addressing the needs of trustworthy operation and stability governance, sequence graph modeling offers a scalable and unified framework that maps call topology, operational signals, and failure semantics into learnable representations, enabling end-to-end extraction of discriminative structural features. This perspective extends beyond failure classification, providing a general foundation for system health assessment, risk prediction, service dependency optimization, and elastic control design. As cloud-native systems continue to grow in scale and complexity, with rapidly evolving services and increasingly diverse observability data, time-series graph–based modeling is poised to become a critical bridge between monitoring data and stability-oriented decision-making, thereby improving system reliability and operational efficiency. Looking ahead, future work can focus on enhancing adaptability to dynamic graph changes driven by scaling and deployment updates, strengthening multi-source semantic alignment across metrics, logs, and traces, integrating online learning for continuous model evolution, and more tightly coupling system representations with automated operation and maintenance loops to enable more efficient and robust fault management in real-world environments.